\documentclass[iop]{emulateapj}

\shorttitle{The Blazhko effect and additional modes}
\shortauthors{Benk\H{o} \& Szab\'o}

\begin{document}

\title{The Blazhko effect and additional excited modes in RR\,Lyrae stars}

\author{J.~M. Benk\H{o} and R. Szab\'o}

\affil{Konkoly Observatory, MTA CSFK, Konkoly Thege Mikl\'os \'ut 15-17., H-1121 Budapest, Hungary}

\email{benko@konkoly.hu}

\begin{abstract}
Recent photometric space missions, such as CoRoT and {\it Kepler} revealed 
that many RR Lyrae stars pulsate -- beyond their main radial pulsation 
mode -- in low amplitude modes.
Space data seem to indicate a clear trend, namely overtone (RRc) stars and modulated
fundamental (RRab) RR\,Lyrae stars ubiquitously show additional modes, while non-Blazhko RRab stars never do. 
Two {\it Kepler} stars (V350\,Lyr and KIC\,7021124), however, 
apparently seemed to break this rule: they were classified 
as non-Blazhko RRab stars showing additional modes.
We processed {\it Kepler} pixel photometric data of these 
stars. We detected small amplitude, but significant 
Blazhko effect for both stars by using the resulted light curves and O$-$C diagrams.
This finding strengthens the apparent connection between the Blazhko effect and
the excitation of additional modes.  In addition, it yields a potential tool for
detecting Blazhko stars through the additional frequency patterns even if
we have only short but accurate time series observations. 
V350 Lyr shows the smallest amplitude multiperiodic Blazhko effect ever detected. 
\end{abstract}


\section{Introduction}

During the past decades RR\,Lyrae stars were regarded
as useful tools for measuring cosmic distances, but otherwise   
rather boring stars. They pulsate radially,
and the mechanism of this pulsation assumed to be well-known.
A few phenomena, however, challenges this simplistic view. 
One of them is the Blazhko effect \citep{Bla07}, 
a periodic amplitude and/or phase variation of the light
curves which is shown by about 50 percent of the RRab stars
and there is no generally accepted physical explanation 
(see e.g \citealt{Szabo14IAUS} for a present review).

The interest in RR\,Lyrae stars has been dramatically increased
through the discoveries of photometric space missions CoRoT \citep{Baglin06}
and {\it Kepler} \citep{Borucki10}.
These objects are either characterized by completely new phenomena, like 
period-doubling \citep{Kolenberg10, Szabo10} and
low amplitude (potentially non-radial) pulsations  
\citep{Chadid10, Benko10}, or phenomena which proved to
be more frequent in the space-based data compared with the ground-based ones, such as 
multiperiodic or irregular Blazhko effects \citep{Guggenberger12, Benko14}.

The observation of \object{AQ\,Leo} by the MOST satellite \citep{Gruberbauer07} was 
the first detailed photometric spaceborne observation which was taken on an RR\,Lyrae star. 
AQ\,Leo is a double-mode pulsating (RRd) star: it pulsates in its fundamental 
and first radial overtone mode simultaneously. 
The frequency analysis of AQ\,Leo showed that its Fourier spectrum contains
an additional frequency and its harmonic beyond the expected frequencies 
of the two radial modes and their linear combinations. 
After analyzing the CoRoT and {\it Kepler} data it turned out that all 
RRc and RRd stars exhibit
 such extra modes \citep{Chadid12, Szabo14, Moskalik15}. A common
property of these additional modes is their period ratios  
with the radial overtone periods which is about 0.61-0.62. The proximity
of these numbers to the reciprocal of the famous golden ratio 
(1.618033\dots) inspired interesting speculations \citep{Linder15}
about the pulsation dynamics. Lately these 
additional modes have been found in ground-based data, as well \citep{Jurcsik15},
what is more, a new group has been identified, where the period ratio 
is about 0.686 instead of 0.618 \citep{OGLE14}.

Interestingly, these frequencies have never been detected in any RRab stars,
but other low amplitude additional frequencies do appear. Some of them
are the half-integer frequencies (HIFs = $1/2f_0$, $3/2f_0$, $\dots$,), where $f_0$
is the frequency of the radial fundamental mode. These are connected to
the period doubling effect. The mostly accepted explanation of this effect is 
a 9:2 resonance between the fundamental and the 9th radial overtone,
a so-called strange mode \citep{Kollath11}. The physical resonance 
destabilizes the fundamental fixed point corresponding to the fundamental mode,
giving rise to a period doubled dynamical state characterized by
alternating maxima and minima in the light curve, and HIFs in 
the Fourier spectra. Other low amplitude additional
frequencies seem to be related to the first ($f_1$) or second ($f_2$)
radial overtones \citep{Benko10}. Some theoretical model computations confirmed 
the possibility of triple resonance states where the fundamental, the
first overtone and a strange modes are exited simultaneously \citep{Molnar12}.
Other resonance combinations (e.g. fundamental and second overtone; 
fundamental, first, and second overtone together; etc.)
which are detected in real stars, have not been modelled yet. 

In the model calculations the appearance of the additional modes are 
independent from the Blazhko effect, but observations suggest a
strong correlation. By unifying the published CoRoT and 
{\it Kepler} Blazhko RRab samples \citep{Kolenberg11, Szabo14, Benko14}
we get 22 stars and among them 17 (77\%) show additional frequencies.
If we do the same comparison for non-Blazhko stars \citep{Nemec11, Nemec13, Szabo14} 
we find 2 stars (\object{V350\,Lyr} and \object{KIC\,7021124}) among 25 (8\%) which pulsate 
in additional modes, as well. 

This paper focuses on these two objects, demonstrating that they are not
exceptions in the sense they do show the Blazhko effect, albeit with very small amplitude.

\section{Data}\label{data}

\begin{table*}
\begin{center}
\caption{Sample from the rectified data file of V350\,Lyr}
\label{sample_data_1}  
\footnotesize{
\begin{tabular}{rcrccrr}
\tableline\tableline
No & Time  &   Flux  & Zero point shift
& Scaling factor &  Corrected flux &  Corrected K$_{\mathrm p}$\\  
     &   (BJD$-$2454833)          &    (e$^{-}$s$^{-1}$)         &    (e$^{-}$s$^{-1}$)   &          
& (e$^{-}$s$^{-1}$)    & (mag)  \\ 
\tableline
         1 &   131.51274 &        5431.0 &     0.00 & 1.073 &  5241.81728032 &    0.20251409 \\
         2 &   131.53318 &        5381.9 &     0.00 & 1.073 &  5192.71476411 &    0.21273262 \\
         3 &   131.55361 &        5333.2 &     0.00 & 1.073 &  5144.01224792 &    0.22296381 \\
         4 &   131.57405 &        5279.5 &     0.00 & 1.073 &  5090.30973174 &    0.23435827 \\
         5 &   131.59448 &        5185.8 &     0.00 & 1.073 &  4996.60721555 &    0.25453076 \\
$\cdots$   &    $\cdots$   &   $\cdots$   &           $\cdots$   &      $\cdots$   &    $\cdots$   &    $\cdots$   \\
\tableline                                               
\end{tabular}
}
\tablecomments{The first five data lines from the file of V350\,Lyr ({\tt table\_kplr009508655.tailor-made.dat}).
The columns contain serial numbers, baricentric Julian dates, flux extracted from the tailor-made 
aperture, zero point shifts (0.0 = no shifting), scaling factors (1.0 = no scaling), 
stitched (shifted, scaled and trend filtered) flux and
their transformation into the K$_{\mathrm p}$ magnitude scale, respectively. See the text for the details.   
}
\end{center}                                            
\end{table*} 
\begin{table*}
\begin{center}
\caption{Sample from the rectified data file of KIC\,7021124}
\label{sample_data_2}
\footnotesize{
\begin{tabular}{rcrccrr}
\tableline\tableline
No & Time  &   Flux  & Zero point shift
& Scaling factor &  Corrected flux &  Corrected K$_{\mathrm p}$\\
     &   (BJD$-$2454833)          &    (e$^{-}$s$^{-1}$)         &    (e$^{-}$s$^{-1}$)   &
& (e$^{-}$s$^{-1}$)    & (mag)  \\
\tableline
         1 &   131.51264 &       19185.6 &   500.00 & 0.946 & 14954.71597491 &   $-$0.16944681  \\
         2 &   131.53307 &       18645.9 &   500.00 & 0.946 & 14415.00956331 &   $-$0.12953873  \\
         3 &   131.55351 &       18174.7 &   500.00 & 0.946 & 13943.80315174 &   $-$0.09345450  \\
         4 &   131.57394 &       17738.7 &   500.00 & 0.946 & 13507.79674020 &   $-$0.05896268  \\
         5 &   131.59437 &       17377.8 &   500.00 & 0.946 & 13146.89032863 &   $-$0.02955899  \\
$\cdots$   &    $\cdots$   &   $\cdots$   &           $\cdots$   &      $\cdots$   &    $\cdots$   &    $\cdots$   \\
\tableline                                               
\end{tabular}
}
\tablecomments{The first five data lines from the file of KIC\,7021124 ({\tt table\_kplr007021124.tailor-made.dat}).
The meaning of the columns are the same as in Table~\ref{sample_data_1}.
}
\end{center}                                            
\end{table*}

We present and analyze those two stars (V350\,Lyr and KIC\,7021124),  
where additional small amplitude modes were discovered, but the Blazhko 
effect has not been detected previously. 
Up to now more than a thousand papers based on {\it Kepler} data
have been published, so the basic features of the mission
are widely known. We refer to \cite{Koch10} and \cite{Jen10a, Jen10b}
for detailed description of the main characteristics of the telescope and
the data. All the technical details are published in the following handbooks: \citet{KIH,DPH,KDCH}.

The {\it Kepler} photometry of the non-Blazhko RR\,Lyrae stars was
studied first by \citet{Nemec11} on the basis of  
the commissioning phase and the first five quarters (Q0-Q5).
Furthermore, \citet{Nemec13}, along with the ground-based spectroscopic
observations, also published  new results from the {\it Kepler} photometry 
of the quarters Q0-Q11.  
The present paper uses the complete  (Q0-Q17) long cadence (LC) {\it Kepler} observations. 

The {\it Kepler} data are publicly available\footnote{via MAST:
\url{https://archive.stsci.edu/kepler/data\_search/search.php}}
 in two forms: 
light curve (SAP, PDC) and target pixel files.
The pixel data require more work to extract precise photometry, 
but for RR Lyrae stars which have relatively large amplitudes the pixel data 
should be more reliable \citep{Benko14}.

Briefly, the predefined apertures of the light curve data are in many cases too small 
(that is, they contain too few pixels), so some fraction of the stellar flux is lost.
This flux loss is typically time dependent and can cause instrumental trends and amplitude changes.
To minimize such effects, we processed the pixel data of the non-Blazhko 
RR\,Lyrae stars in the same manner as we did for the Blazhko stars: (1) we defined
a tailor-made aperture for each star and observing quarter, 
then (2) we extracted the flux. (3) The flux data of the different quarters of a star 
were stitched together by scaling and/or shifting. (4) Finally, we removed
the likely instrumental trends and transformed the flux values to the magnitude scale.
The interested reader is referred to \citet{Benko14} for 
details. Tables~\ref{sample_data_1} and \ref{sample_data_2} show 
excerpts from the processed data files as
an example\footnote{All rectified light curves are available at
\url{http://www.konkoly.hu/KIK/data.html}.}. 
Both former studies of non-modulated {\it Kepler} RR\,Lyrae stars 
\citep{Nemec11,Nemec13} used the {\it Kepler} light curves. 
This letter is the first, where the pixel data of non-Blazhko stars are 
used.

\section{Analysis and Results}

Our main tools are the Fourier analysis 
realized by the {\sc MuFrAn} program package \citep{Mufran}
and the `observed minus calculated' (O$-$C) diagram method 
(see e.g. \citealt{Sterken05}).
The O$-$C diagram is calculated from the maxima time of the light curve. 
The details of the analysis are summarized in \cite{Benko14}.  
Throughout this paper the numerical values (frequencies, 
amplitudes, etc.) are written with the significant number 
of digits plus one digit.

\subsection{V350\,Lyr = KIC\,9508655}

\begin{figure}
\includegraphics[width=8.5cm]{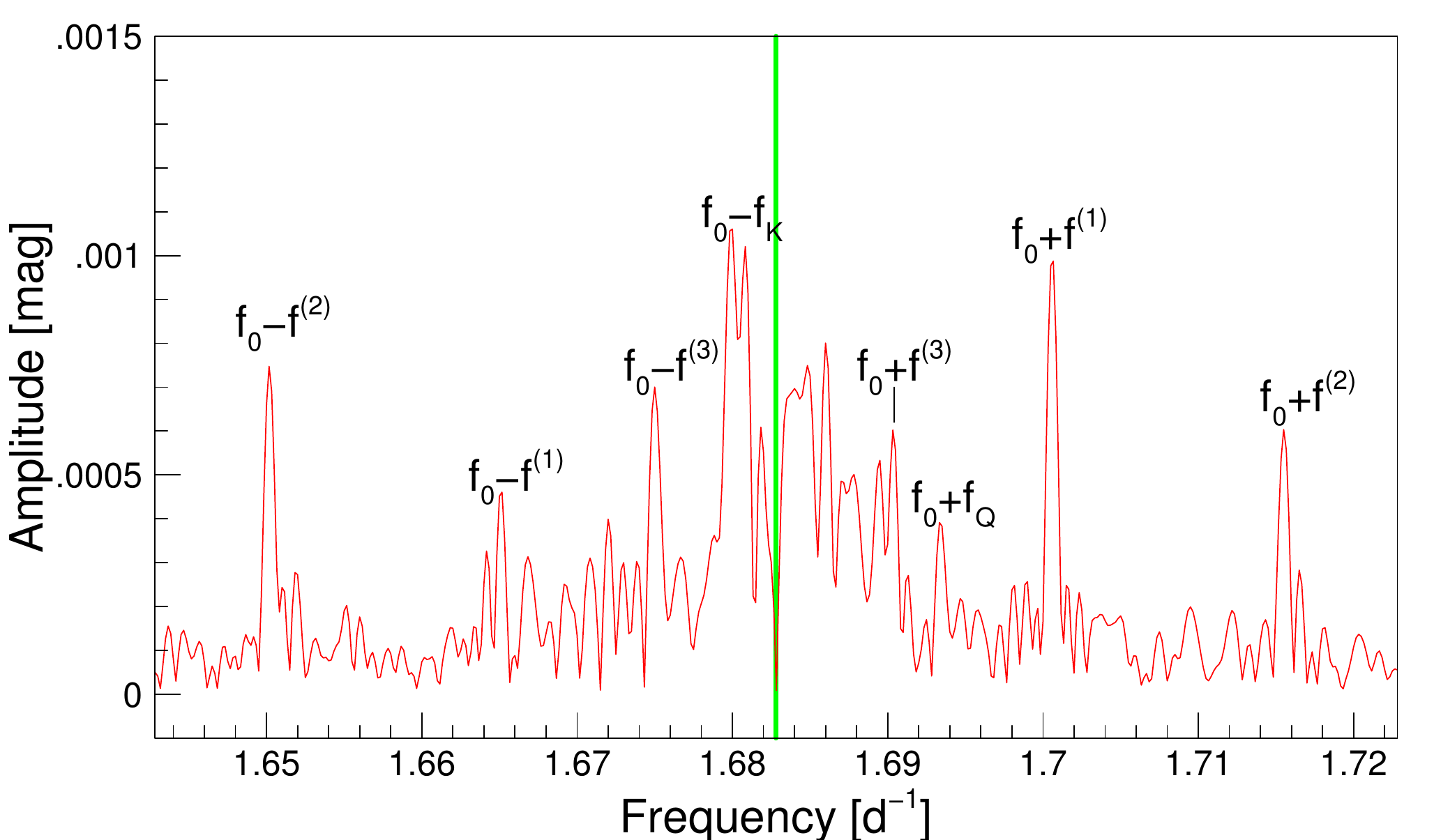}

\includegraphics[width=8.5cm]{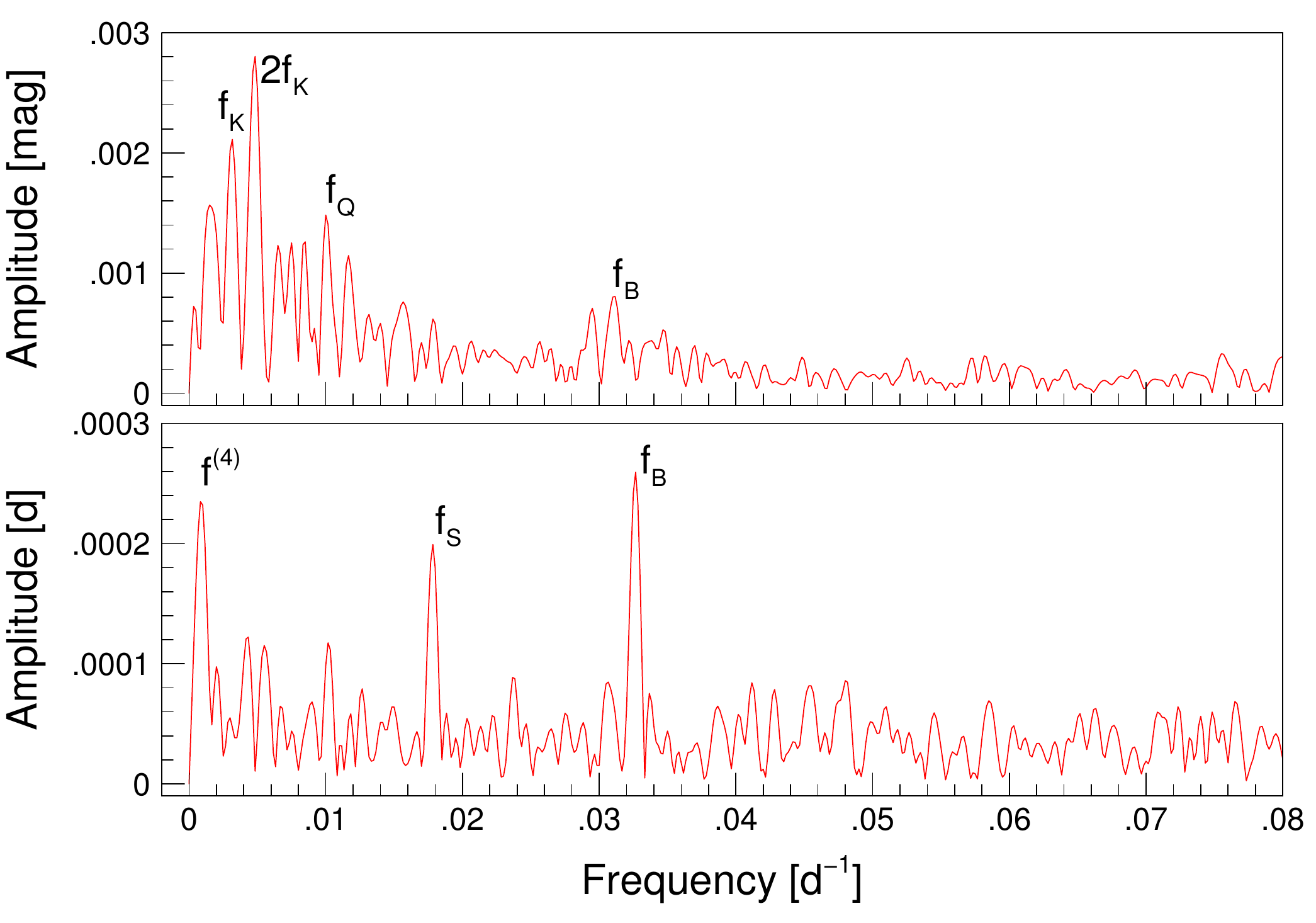}
\caption[]{
Side peaks around the main pulsation frequency of V350\,Lyr after a 
pre-whitening step (top). The green vertical line shows the position of the
pre-whitened main pulsation frequency $f_0$.
Low frequency range of the Fourier spectra of V350\,Lyr light curve (middle) 
and O$-$C diagram (bottom). See text for more details.
} \label{V350_Lyr_fr}
\end{figure}
The observational history of this star is rather short.
\citet{Hoffmeister66} discovered and classified it as an RR\,Lyrae
variable star, giving two maximum times. 
\citet{Galkina85} determined some basic photometric parameters 
(epoch, period, maximum, minimum values, and amplitude) from
their photographic observations. They assumed a period variation on 
longer time scale, because they could not find a common period for
their maxima and Hoffmeister's ones.

\begin{figure*}[t!!!]
\includegraphics[width=17cm]{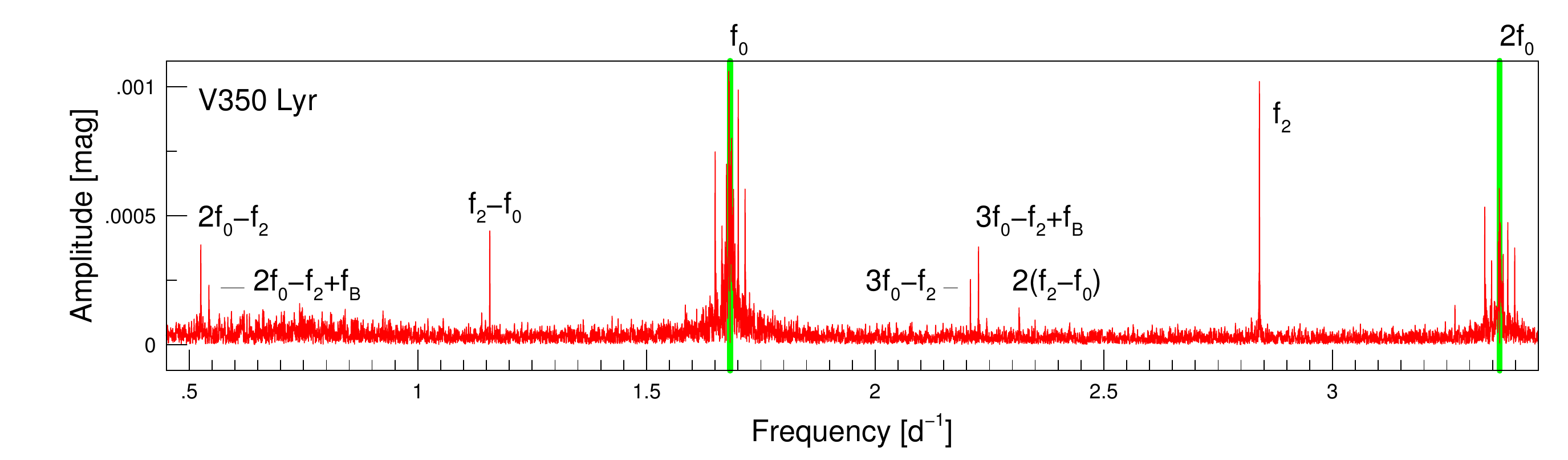}
\includegraphics[width=17cm]{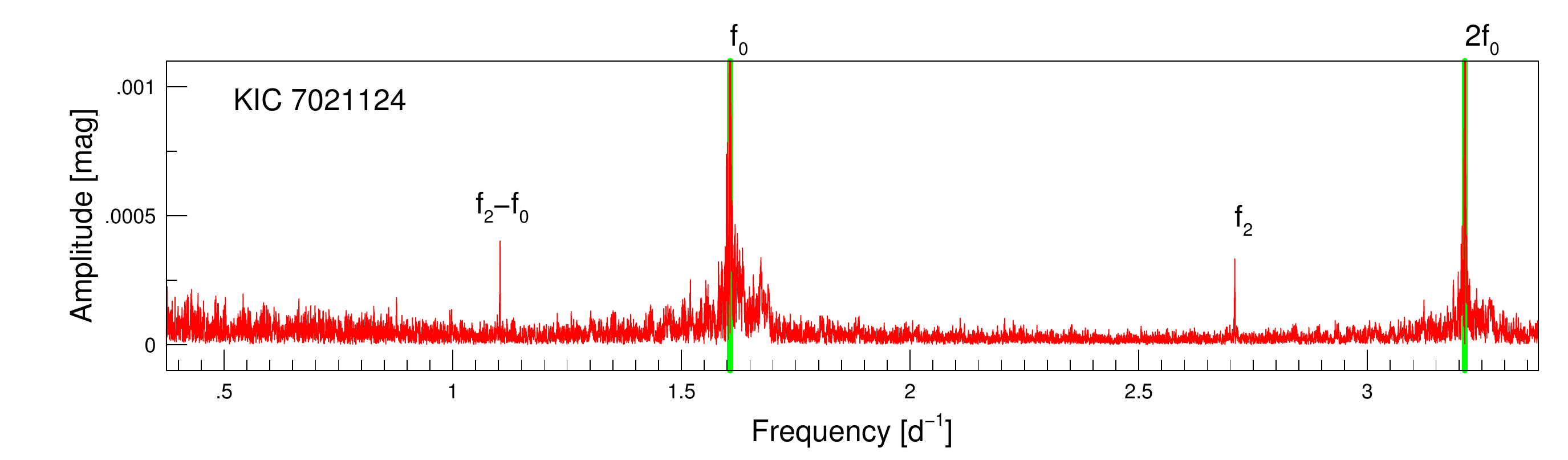}
\caption[]{
Additional peak identifications in the 
the pre-whitened Fourier spectrum of V350\,Lyr (top) and 
KIC\,7021124 (bottom) light curves.
The green vertical lines show the positions of the pre-whitened
frequencies $f_0$ and $2f_0$. 
} \label{add}
\end{figure*}   
After a long hiatus, {\it Kepler} entered the scene.
V350\,Lyr was classified by \citet{Benko10} as a non-Blazhko star on 
the basis of 138~d (Q1-Q2) {\it Kepler} observations though the 
possibility of a small amplitude Blazhko effect below the detection 
limit was also noted, because the residual spectrum after pre-whitening
with the main pulsation frequency $f_0=1.682814$~d$^{-1}$ and its harmonics,
showed a bunch of small amplitude peaks around each pre-whitened frequency. 
  An additional frequency at $f=2.84019$~d$^{-1}$ and its 
linear combination with the main pulsation frequency was detected 
as the most interesting feature of this star. 
At that time V350\,Lyr was though to be the first 
and the only example for a non-Blazhko RRab star pulsating in an additional mode.
The frequency was identified with the second radial overtone $f=f_2$.
\citet{Nemec11} confirmed both findings (non-Blazhko behavior and the excited additional
mode) using the {\it Kepler} light curve data from Q1-Q5.
In the following we demonstrate that in contrast with earlier results based on
shorter {\it Kepler} data, V350\,Lyr is indeed a Blazhko-modulated star.

To do this, now we turn to the analysis of the latest available {\it Kepler} data.
The Fourier spectrum of the rectified light curve based on Q1-Q17 pixel data is dominated by
the main pulsation frequency ($f_0=1.682828$~d$^{-1}$) and its harmonics
($kf_0$, $k=1,2,\dots,$). 
Thirteen  harmonics can be detected up to the Nyquist frequency (24.5~d$^{-1}$).
By pre-whitening the light curve with these dominant frequencies, we found
multiplet structures around their positions (top panel in Fig.~\ref{V350_Lyr_fr}):
$kf_0\pm f^{(i)}$, where $k=1,2,\dots,$ and $i=1, 2$ or 3.
Assuming that these multiplets are combinations of some modulation side peaks, 
 we can calculate the individual modulation frequencies.   
The averaged differences of the $f^{(i)}$ frequencies from the two side peaks around $f_0$ 
are $f^{(1)}=0.01773$~d$^{-1}$, $f^{(2)}=0.03265$~d$^{-1}$, and 
$f^{(3)}=0.00772$~d$^{-1}$, respectively. Many stars in our 
sample show a frequency around 0.008~d$^{-1}$, so these frequencies together
with $f^{(3)}$ must be of instrumental origin. Some additional instrumental peaks 
(e.g. $f_0-f_{\mathrm K}$, $f_0+f_{\mathrm Q}$, see Fig.~\ref{V350_Lyr_fr}) are also detectable. Here
we define the frequencies belonging to the {\it Kepler} year as $f_{\mathrm K}=1/372.5$~d$^{-1}$
and to average the quarter as $f_{\mathrm Q}\approx 1/90$~d$^{-1}$.
We introduce the notation $f^{(2)}=f_{\mathrm B}$ and $f^{(1)}=f_{\mathrm S}$
for the intrinsic modulations, the primary and secondary Blazhko frequencies, respectively.

It is noticeable, that $f_{\mathrm B}$ and $f_{\mathrm S}$ are nearly harmonics:  
$f_{\mathrm B} \approx 2f_{\mathrm S}$. However, the difference
between the exact harmonic and the actual value ($2f_{\mathrm S}-f_{\mathrm B}=0.0028$) 
is significantly higher than the Rayleigh frequency resolution ($\approx 0.0007$~d$^{-1}$). 
This means that we are facing a multiperiodic Blazhko modulation with nearly resonant
frequencies similar to \object{CZ\,Lac}, \object{RZ\,Lyr} \citep{Sodor11, Jurcsik12} 
and numerous cases in the {\it Kepler} Blazhko sample \citep{Benko14}.

The low frequency range of the Fourier spectrum (middle panel in Fig.~\ref{V350_Lyr_fr}) 
is dominated by 
the technical peaks, but 
$f_{\mathrm B}$ ($A^{\mathrm{AM}}(f_{\mathrm B})=0.8$~mmag)\footnote{From now on, 
the upper indices AM and FM distinguish the
amplitude (AM) and frequency modulation (FM) amplitudes, respectively.} 
can also be detected. 
The peak at $f_{\mathrm S}$, however,
is not significant ($A^{\mathrm{AM}}(f_{\mathrm S})=0.6$~mmag).
The highest peak between 
the harmonics is $f_2=2.840182$~d$^{-1}$ (top panel in  Fig.~\ref{add}). 
We also detect numerous linear combination
frequencies (such as 1.157383~d$^{-1}$ $=f_2-f_0$, 0.525460~d$^{-1}$ = $2f_0-f_2$,
2.314693 = $2(f_2-f_0)$, etc.).  
The Fourier spectrum features clearly demonstrate that
V350\,Lyr is a typical Blazhko RR\,Lyrae star. Nevertheless, the LC light curve does not
show evident modulation. The rms of the non-linear fit using the main
frequency and its harmonics is 0.0062~mag. This value is 0.0048~mag for the
non-modulated \object{V1107\,Cyg} which has the closest brightness and period parameters to 
V350\,Lyr in the {\it Kepler} RR\,Lyr sample. This small difference might be explained by the
effect of the additional frequencies of V350\,Lyr.

To clarify the situation we prepared the O$-$C diagram, that tests the
frequency modulation part of the potential Blazhko effect.
The O$-$C diagram itself does not show evident modulation, however, its
Fourier spectrum (bottom panel in Fig.~\ref{V350_Lyr_fr}) contains three 
significant frequencies: $f_{\mathrm B}$, 
$f_{\mathrm S}$, and a third 
one at $f^{(4)}=0.000861$~d$^{-1}$ with the amplitudes 
of $A^{\mathrm{FM}}(f_{\mathrm B})=0.0003$~d, 
$A^{\mathrm{FM}}(f_{\mathrm S})=0.0002$~d, and 
$A(f^{(4)})=0.0003$~d, respectively.
 The latter frequency belongs to a long 
period variation ($P^{(4)}\approx 3.1$~years) which has no signs in the light
curve or the light curve spectrum. At the same time $f^{(3)}$ can not be 
detected here in the O$-$C spectrum, which supports its technical origin.

\begin{figure}
\includegraphics[width=8.5cm]{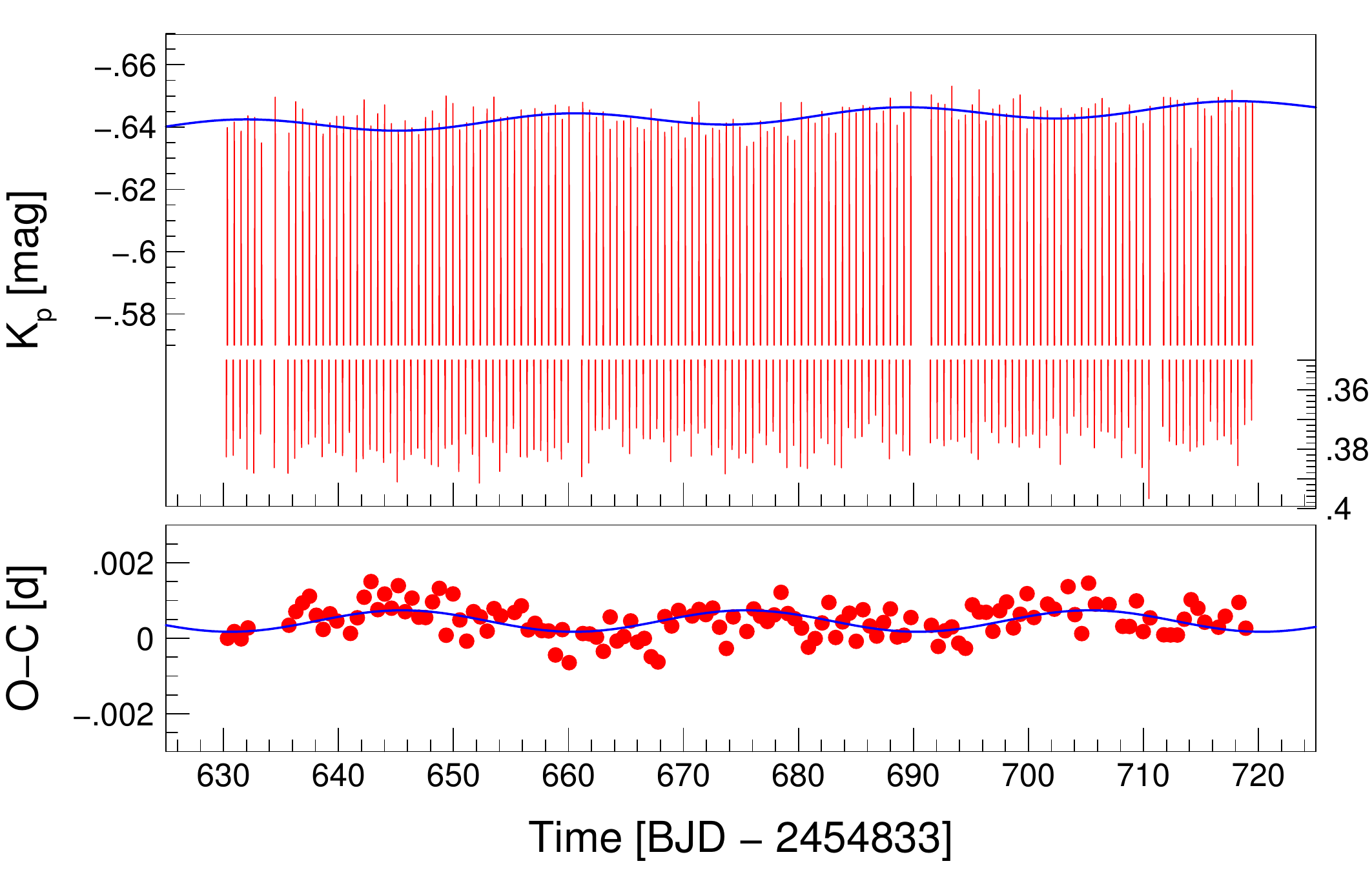}
\caption[]{
(top) Short cadence (SC) light curve of V350\,Lyr from Q7. For clarity the 
middle values of the light curve are not plotted here, 
just the two extrema (minima and maxima).
(bottom) O$-$C diagram of the SC light curve above. The continuous blue lines show
the best fit sinusoid of the parameters $P_{\mathrm B}=28.7$~d,
 $A=0.0025$~mag (for maxima) and $P_{\mathrm B}=29.9$~d,
$A^{\mathrm {FM}}(f_{\mathrm B})=0.00028$~d (for O$-$C values).
} \label{v350_sc}
\end{figure}

V350\,Lyr was observed in SC mode in Q7 and  for one month in Q11.3. 
These observations give us an additional opportunity to check
the Blazhko nature of this star. 
We processed the SC pixel data exactly the same way 
as we did the LC data. The SC light curve indeed shows slight amplitude changes
(top panel in Fig.~\ref{v350_sc}). The magnitude of this variation is less than 0.005~mag.
The O$-$C diagram of the SC data (bottom panel in Fig.~\ref{v350_sc}) shows 
expressed variation with a period of 30~days which can be identified 
with the Blazhko period $P_{\mathrm B}$.

We conclude that V350\,Lyr is a Blazhko star showing (at least)
two modulations with small variation amplitude 
and frequency, and these two modulation frequencies 
are in nearly 1:2 resonance. 
The O$-$C diagram shows $f^{(4)}=0.000861$~d$^{-1}$ beyond the two Blazhko 
frequencies. This frequency could either belong to (1) a third Blazhko modulation
or (2) is the consequence of the light-time effect 
caused by a gravitational bound companion. 
Such O$-$C diagrams were studied by and \citet{Hajdu15} 
and \citet{Guggenberger14} in the case of RR Lyrae stars, recently. (3) Less likely:
of instrumental origin.

\subsection{KIC\,7021124}

\begin{figure}
\includegraphics[width=8.5cm]{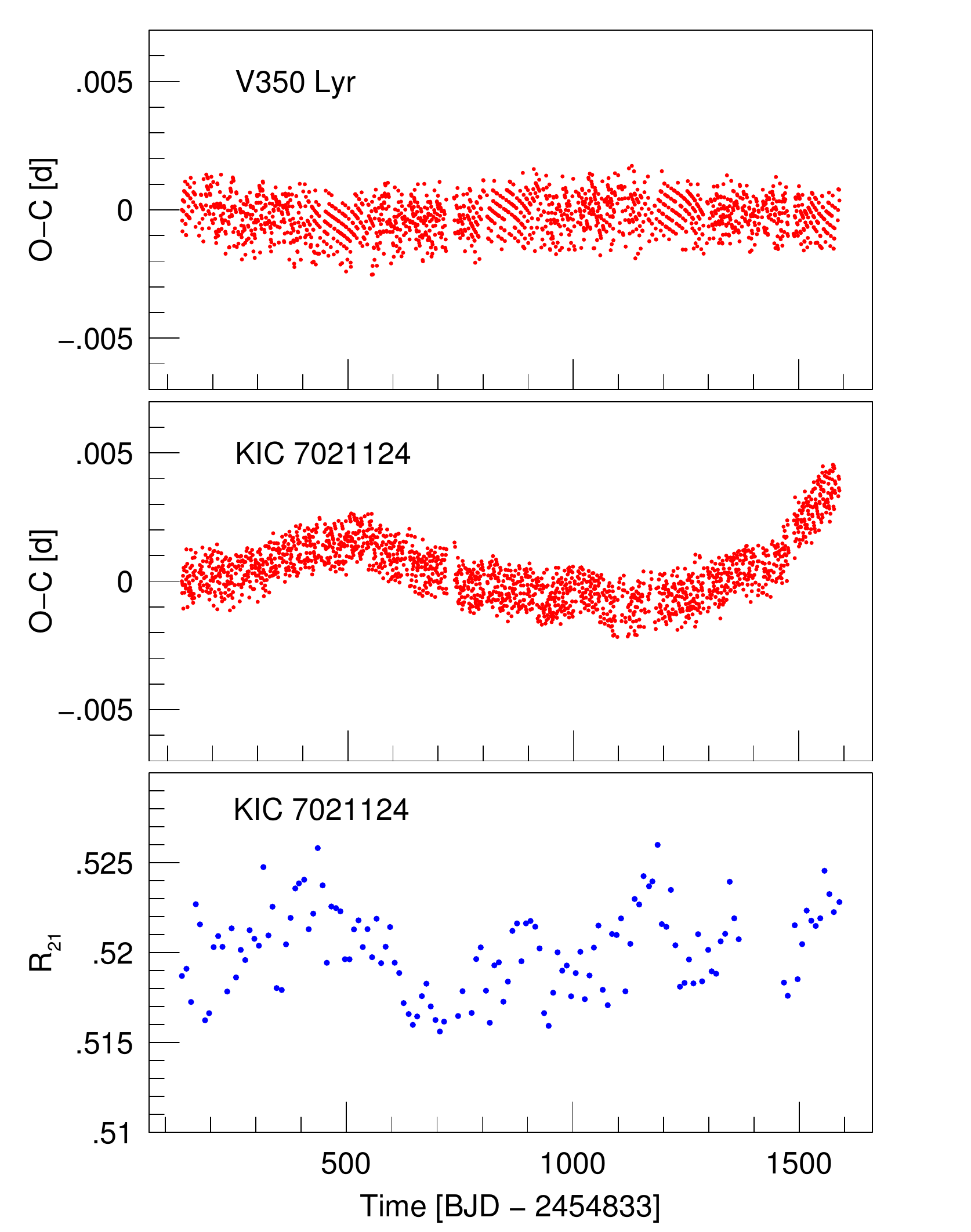}
\caption[]{
O$-$C diagrams of V350\,Lyr (top). O$-$C diagrams of KIC\,7021124 (middle) and
the time dependence of its amplitude ratio $R_{21}$ (bottom).
} \label{o-c_k}
\end{figure}   
The additional mode of KIC\,7021124 was discovered by \citet{Nemec11}
making this star the second non-Blazhko star showing additional mode at that time.
\citet{Nemec11} found this star to be very similar to V350\,Lyr in many aspects
(e.g. mass, luminosity, Fourier parameters).
At that time only Q1 data were available. 
Here we deprive this star of this privileged status, as well.

Using the entire data set from Q1 to Q17 we do not find significant amplitude modulation.
The pre-whitened spectrum does not show distinct side peaks around the
harmonics of the main pulsation period $f_0=1.606474$~d$^{-1}$ 
and the significant peaks in the low frequency range are presumably technical.  
The spectrum contains an additional frequency $f_2=2.70999$~d$^{-1}$ (see bottom 
panel in Fig.~\ref{add}) and
some of its linear combination (e.g. $f_2-f_0$), but generally it is more simple
than the spectrum of V350\,Lyr.

The O$-$C diagram, however, shows a clear period change (middle panel in Fig.~\ref{o-c_k}).
The shape of the O$-$C curve is close to, but not strictly sinusoidal. In the Fourier 
spectrum of this O$-$C diagram there are two significant peaks at 
$f^{(5)}=0.00087$~d$^{-1}$,
($A(f^{(5)})=0.0011$~d) and at $2f^{(5)}=0.0019$~d$^{-1}$, ($A(2f^{(5)})=0.0005$~d). 
Further peaks can not be detected in the higher frequency range.  
These frequencies yield a rough period estimation
of around $\sim 1400$~d, since the variation period -- if it is periodic at all -- 
is comparable to the total observing time. 
The appearance of the overtone indicates the non strictly sinusoidal
nature of the variation. The similar period
of $f^{(4)}$ and $f^{(5)}$ raises the possibility that these variations result from
instrumental effects. Some facts contradict such a scenario.
(1) It is highly unlikely that there are problems with the time measurements of {\it Kepler}. 
(2) The amplitude $A(f^{(5)})=0.0011$~d=1.55~min 
is much higher than that of the longest period of V350\,Lyr 
($A(f^{(4)})=0.0003$~d=0.4~min).
(3) The phase of these two similar time-scale variations are also different 
(cf. top and middle panels in Fig.~\ref{o-c_k}). 
All in all, both variations described by $f^{(4)}$ and $f^{(5)}$ seem to be real. 
The question of their nature however, remains.

In the case of V350\,Lyr we have already mentioned some possible scenarios.
 The frequency modulation due to a 
long period Blazhko effect would be an adequate explanation for KIC\,7021124 as well,
because  Blazhko cycles of long characteristic time scales are known \citep{Soszynski11}. 
The only problem is the lack of the amplitude modulation. It is known, however, that if we
characterize the light curve with the Fourier parameters defined by \cite{Simon82}, 
the amplitude ratio $R_{21}=A(2f_0)/A(f_0)$ is very sensitive to the amplitude
changes and is highly unaffected by technical problems.
The bottom panel of Fig.~\ref{o-c_k} shows the variation of $R_{21}$ in time.
The diagram was constructed with the {\sc Period04} program \citep{Period04}. 
The $R_{21}$ parameter shows a long-term variation, very similar to that of the O$-$C values.
The phase of the long time-scale variation is correlated with the O$-$C diagram.

Simultaneous amplitude and frequency variation
constitutes a strong evidence for the presence of the Blazhko effect in this object.
Its small amplitude and long cycle prevented its detection until now.

\section{Conclusions}

We extracted and presented the time series of two RR Lyrae stars (V350\,Lyr and KIC\,7021124)
-- which were formerly known as un-modulated ones  
showing additional excited small amplitude periodicites.
We used the {\it Kepler} pixel data and our tailor-made apertures to
minimize flux loss. The flux curves 
were scaled, shifted and de-trended in the same way as was done for the Blazhko stars
\citep{Benko14}.

The study of these two stars  
resulted in evidence for the Blazhko behavior in both cases.
In the case of V350\,Lyr we demonstrated that this star shows
simultaneous amplitude and frequency modulations 
 with two small amplitude frequencies with nearly 1:2 ratio.    
KIC\,7021124 is also a Blazhko star
with extremely low amplitude modulation at about the {\it Kepler}
detection limit, featuring a significant long-period frequency
modulation. 

The smallest known Blazhko AM amplitude so far has been 0.6~mmag 
for V838\,Cyg \citep{Nemec13}. The secondary AM modulation
amplitude of V350\,Lyr has the same amplitude, however, this object shows 
multiple modulations with the smallest amplitude components (0.6 and 0.8~mmag) ever found.

KIC\,7021124 is also a record holder: it has by far the
longest Blazhko period with such a small amplitude. A possible trend
was reported between the Blazhko period and the amplitude of the 
AM parts of the effect (see in Fig.~9 in
\citealt{Benko14}). KIC\,7021124 seems to diverge from this 
trend. This points out that the trend might partially be a sampling effect:
long period and small amplitude modulations can hardly be detected even from space.

We mention that these two stars are the most metal poor stars in 
the {\it Kepler} Blazhko sample: V350\,Lyr has [Fe/H]=$-1.83$~dex,
and KIC\,7021124 has [Fe/H]=$-2.18$~dex \citep{Nemec13}. 
The more metal poor RR\,Lyrae stars in the sample (NR\,Lyr, FN\,Lyr, NQ\,Lyr) 
are all non-Blazhko stars. Is the amplitude of the AM part of the Blazhko effect related 
to the metallicity? A direct metallicity -- AM amplitude relation can be ruled
out because e.g. V838\,Cyg has also small AM amplitude but it is 
the most metal rich ([Fe/H]=$-1.01$~dex) among the {\it Kepler} Blazhko stars.
If we complement the {\it Kepler} Blazhko sample with V350\,Lyr and KIC\,7021124
we find that the metallicities of the non-Blazhko stars 
distribute over a wider range (between $-2.54$ and $-0.05$~dex) than that 
of the Blazhko stars ($-2.18$ and $-1.01$). What is even more interesting: 
both the extremely metal rich and metal poor Blazhko stars exhibit extremely low
AM amplitude.

On the basis of the CoRoT and {\it Kepler} Blazhko samples -- and  
since both objects studied here proved to be Blazhko stars --
 we can provide a {\it strict rule for RRab stars:
the additional modes appear only in the presence of the Blazhko effect}.
In this case we were able to deduce the Blazhko nature for those stars
where the Blazhko cycle is long (much longer than the observed time span),
 but some excited additional modes are evident.
This situation will be common in the relatively short observing runs
of K2 \citep{Howell14} and TESS \citep{Ricker15}. 

\acknowledgments

Funding for the {\it Kepler} Discovery mission is
provided by NASA's Science Mission Directorate. The authors thank the 
entire {\it Kepler} team without whom these results would not be possible.
BJM would like to thank to E. Plachy and J.M. Nemec for their valuable comments.
This project has been supported by the `Lend\"ulet-2009 Young Researchers' Program
of the Hungarian Academy of Sciences and through the ESA PECS Contract No. 4000110889/14/NL/NDe.
The research leading to these results has received funding from the European Community's
Seventh Framework Programme (FP7/2007-2013) under grant agreements no.~269194 (IRSES/ASK)
and no.~312844 (SPACEINN).

\end{document}